\documentclass[aps,pra,groupedaddress,twocolumn]{revtex4}
\usepackage{mathtools}
\usepackage[colorlinks]{hyperref}
\newcommand{\ket}[1]{\vert #1 \rangle}

\begin{document}

\title{Strongly correlated states of light in chiral chains of three-level quantum emitters}

\author{Ole Aae Iversen}
\author{Thomas Pohl}
\affiliation{Department of Physics and Astronomy, Aarhus University, Ny Munkegade 120, DK-8000 Aarhus C, Denmark}

\date{\today}

\begin{abstract}
We study the correlated transport of photons through a chain of three-level emitters that are coupled chirally to a photonic mode of a waveguide. It is found that this system can transfer a classical input into a strongly correlated state of light in a unitary manner, i.e. without the necessity of nonlinear photon losses. In particular, we shows that the collective interaction with the emitter ensemble leads to the emergence of highly antibunched light with long-range correlations upon crossing a critical length of the chain. By operating close to conditions of  electromagnetically induced transparency of the three-level medium, the high degree of antibunching and photon transmission can be maintained in the presence of moderate losses. These features, combined with the robustness against number fluctuations, suggest a promising mechanism for single-photon generation and may open the door to exploring correlated quantum many-body states of light. 
\end{abstract}


\maketitle

The rapidly expanding capabilities for single-particle control of quantum many-body systems has opened up new research into nonlinear optics at the ultimate quantum level \cite{Chang2014,chiral_quantum_optics} by exploiting cooperative and collective phenomena in large assemblies of quantum emitters. This includes gases of atoms with strongly interacting Rydberg states \cite{Pritchard2010,Peyronel2012,Murray2016a} that can block the light-matter coupling of multiple nearby photons, which gives rise to effective photon-photon interactions \cite{Thompson2017,Tiarks2019,Liang2018,Stiesdal2018,stiesdal2020}, and may also be achieved by placing ground state atoms at extreme distances well below the optical wavelength \cite{cidrim2020,williamson2020}. One can also use structured arrangements of quantum emitters in one \cite{Garcia2017} and two dimensions \cite{Bettles2016,Shahmoon2017,schuler2020} for an increased light-matter coupling and control of cooperative effects \cite{Meir14,Sutherland16}, as observed in very recent experiments \cite{rui2020,glicenstein2020}. Hereby, optical interfaces, such as resonators or waveguides, offer an efficient means for mode-selective coupling to enhance the intrinsic nonlinearity of single emitters \cite{Welte2018,Sayrin2015}, induce interesting photon scattering dynamics and correlated transport \cite{sf07,sf07-prl,zgb10,roy10,fks10,Pletyukhov_2012,Ringel_2014,shi_chang_cirac_15,fb15,song_et_al_17,song_et_al_18,Kumlin18,wang20}, and to employ such effects to manipulate few-photon states of light \cite{Javadi2015,Goban2014,Tiecke2014,Prasad_2019}.

Remarkably, this has been possible \cite{Prasad_2019} in waveguide-coupled ensembles of two-level atoms without mutual interactions, regular arrangements, or sub-wavelength spacings of the emitters. Nevertheless, strong photon correlations were found to emerge in this setting under unidirectional propagation along an optical nanofiber purely from the interplay of interference and dissipative photon loss \cite{Mahmoodian_2018}. At a specific ensemble size this competition can entirely inhibit the simultaneous transmission of two photons and therefore suggests a promising mechanism for single-photon generation, limited only by the achievable control of atom number fluctuations and the optical transmission in the presence of the required photon losses. 

\begin{figure}[t!]
		\includegraphics[width=0.95\columnwidth]{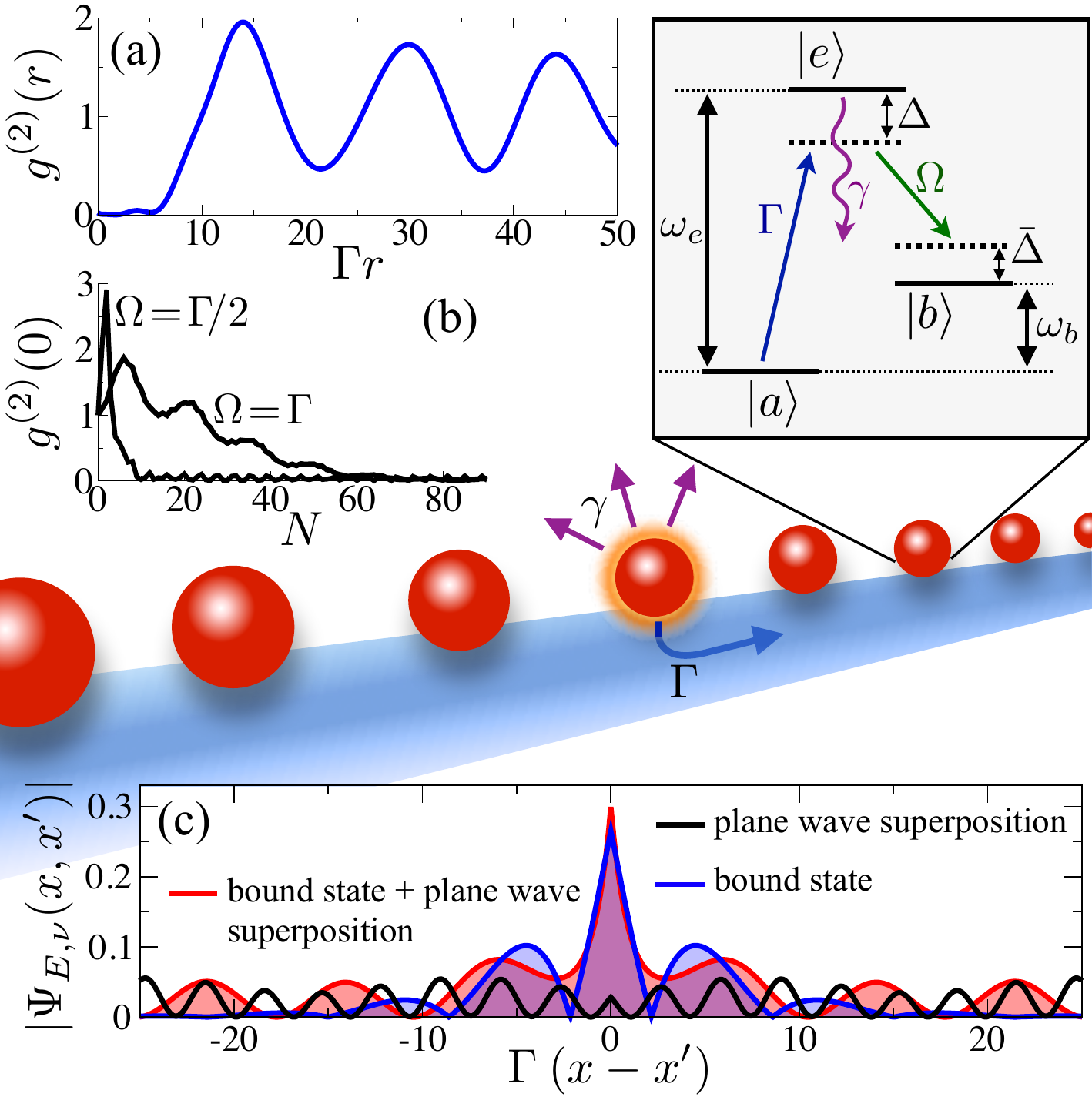}
\caption{A chain of 3-level emitters is chirally coupled to a photonic mode of a waveguide. The guided photons drive a transition between a stable state $\ket{a}$ and an excited state $\ket{e}$, which is turn is coupled to another stable state $\ket{b}$. The corresponding coupling strength, level energies and frequency detunings are indicated in the depicted level diagram. The collective coupling to a chain of such three-level systems can give rise to extended-range correlation, as indicated by the two-photon correlation function $g^{(2)}(x)$ of transmitted photons in panel (a). Strongly antibunched photons with $g^{(2)}(x)\approx0$ can emerge upon crossing a critical chain length $N$, as shown in panel (b) for different indicated control field Rabi frequencies $\Omega$. The photon output is determined by the eigenstates of the underlying scattering matrix, and we find three types of such states, which are illustrated in panel (c) and discussed in more detail in the main text. We have used $\Delta = \Gamma/4$ and $\bar{\Delta} = -\Gamma/4$ in all calculations and $\Omega = \Gamma/2$ for (a) and (c). \label{fig:system}}
\end{figure}

Here, we investigate correlated photon transport through a chiral chain of $\Lambda$-type three-level emitters [Fig. \ref{fig:system}] and reveal an efficient mechanism for strong photon antibunching that does neither require dissipation nor a precise tuning of the number of emitters. Optical coupling to a third level gives rise to a new type of photon-scattering state [Fig. \ref{fig:system}(c)] and facilitates the formation of strong and long-range photon correlations [Fig. \ref{fig:system}(a)] without a dissipative nonlinearity, but via a coherent redistribution of photons as they propagate through the chain of otherwise noninteracting quantum emitters. Remarkably, this can lead to a steady growth of photonic antibunching as the length of the chain is increased and eventually yields highly antibunched light beyond a critical number of emitters [Fig. \ref{fig:system}(b)]. Our scheme therefore offers a new approach to generate strongly antibunched light that does not rely on the otherwise challenging requirements of strong atomic interactions, accurate control of particle numbers, or regular sub-wavelength arrangements of quantum emitters. For example, this suggests a promising route to realizing bright sources of single photons and may open up explorations of self-organization phenomena that may even include the spontaneous emergence of regular trains of single-photon pulses.

The considered setting consists of a chiral single-mode waveguide \cite{chiral_quantum_optics} aligned with a chain of $N$ three-level emitters at positions $x_j$ along the waveguide ($j=1,\dots,N$). The emitters feature three internal states $|a_j\rangle$, $|e_j\rangle$, and $|b_j\rangle$ with associated excited state energies $\omega_e$ and $\omega_b$, as illustrated in Fig. \ref{fig:system}. Hereby, $|a\rangle$ and $|b\rangle$ represent stable states and the excited state $|e\rangle$ decays into the waveguide mode with a rate $\Gamma$ and into non-guided modes of its surrounding with a rate $\gamma$. The waveguide chirality implies that only the right-propagating waveguide mode couples to the $|a\rangle-|b\rangle$ transition of the emitters \cite{Petersen_2014}, as indicated in Fig. \ref{fig:system}. Photons in that mode are created by the bosonic field operator $\hat{c}^\dagger(x)$ and can transfer an emitter into the excited state $|e\rangle$, from where it is coupled to another stable state $|b\rangle$ by a classical laser field with a frequency $\bar{\omega}$ and Rabi coupling $\Omega$. The corresponding Hamiltonian can be written as 
\begin{eqnarray}\label{eq:H}
		\hat{H}&=& -i\int {\rm d}x \hat{c}^{\dagger}(x)\partial_{x} \hat{c}(x)\nonumber\\
		&& + \sum_j\left[\omega_{e} - i\frac{\gamma}{2}\right]\hat{\sigma}_{ee}^{(j)} + (\omega_b + \bar{\omega})\hat{\sigma}_{bb}^{(j)}, \\
		&&+ \sum_j\bigg[\int \sqrt{\Gamma}\delta(x - x_j) \hat{c}(x)\hat{\sigma}_{ea}^{(j)}{\rm d}x + \Omega \hat{\sigma}_{be}^{(j)} + \rm{h.c.}\bigg], \nonumber
\end{eqnarray}
where $\hat{\sigma}_{\alpha\beta}^{(i)}=|\alpha_i\rangle\langle\beta_i|$, and the chosen units are such that $\hbar=1$ and velocities are scaled by the speed of light in the waveguide. We will study the propagation of two photons with well-defined identical incident energies of $\omega$. The total incident energy $E=2\omega$ can thus be parametrized in terms of the single-photon detuning $\Delta = \omega_e - \omega$ and the two-photon detuning $\bar{\Delta} = \omega_b - (\omega - \bar{\omega})$ of the three-level system [cf. Fig. \ref{fig:system}].

For a sufficiently weak coherent input field the emerging correlations in the light field are dominated by its two-photon component, such that one can restrict the analysis to the simultaneous propagation of two photons across the chain. Generally, the solution to this problem can be constructed from the eigenstates of two photons interacting with a single emitter \cite{Mahmoodian_2018}. The corresponding state
\begin{eqnarray}
		|\phi\rangle &=& \frac{1}{\sqrt{2}}\int {\rm d}x {\rm d}x^\prime \psi(x,x^\prime)\hat{c}^{\dagger}(x)\hat{c}^{\dagger}(x^\prime)|0,a\rangle \nonumber\\
			&& + \int {\rm d}x\left[e(x)\hat{\sigma}_{ea} + b(x)\hat{\sigma}_{ba}\right]\hat{c}^{\dagger}(x) |0,a\rangle \label{eq:scattering-state}
\end{eqnarray}
can be decomposed into the two-photon amplitude $\psi(x,x^\prime)$ and the amplitudes $e(x)$ and $b(x)$ to find one photon at position $x$, while the other photon has been absorbed to excite the emitter to the state $|e\rangle$ or $|b\rangle$, respectively.
 Here $|0,a\rangle$ denotes the state with zero photons and the emitter in the ground state $|a\rangle$. One obtains a set of coupled Schr\"odinger equations for the amplitudes in Eq. (\ref{eq:scattering-state}), which have been analyzed previously for the linear regime of single photons \cite{yan_et_al_2018,ws10} and scattering off single atoms \cite{roy11,zgb12,rb14,fb16}.

\begin{figure*}
		\includegraphics[width=\textwidth]{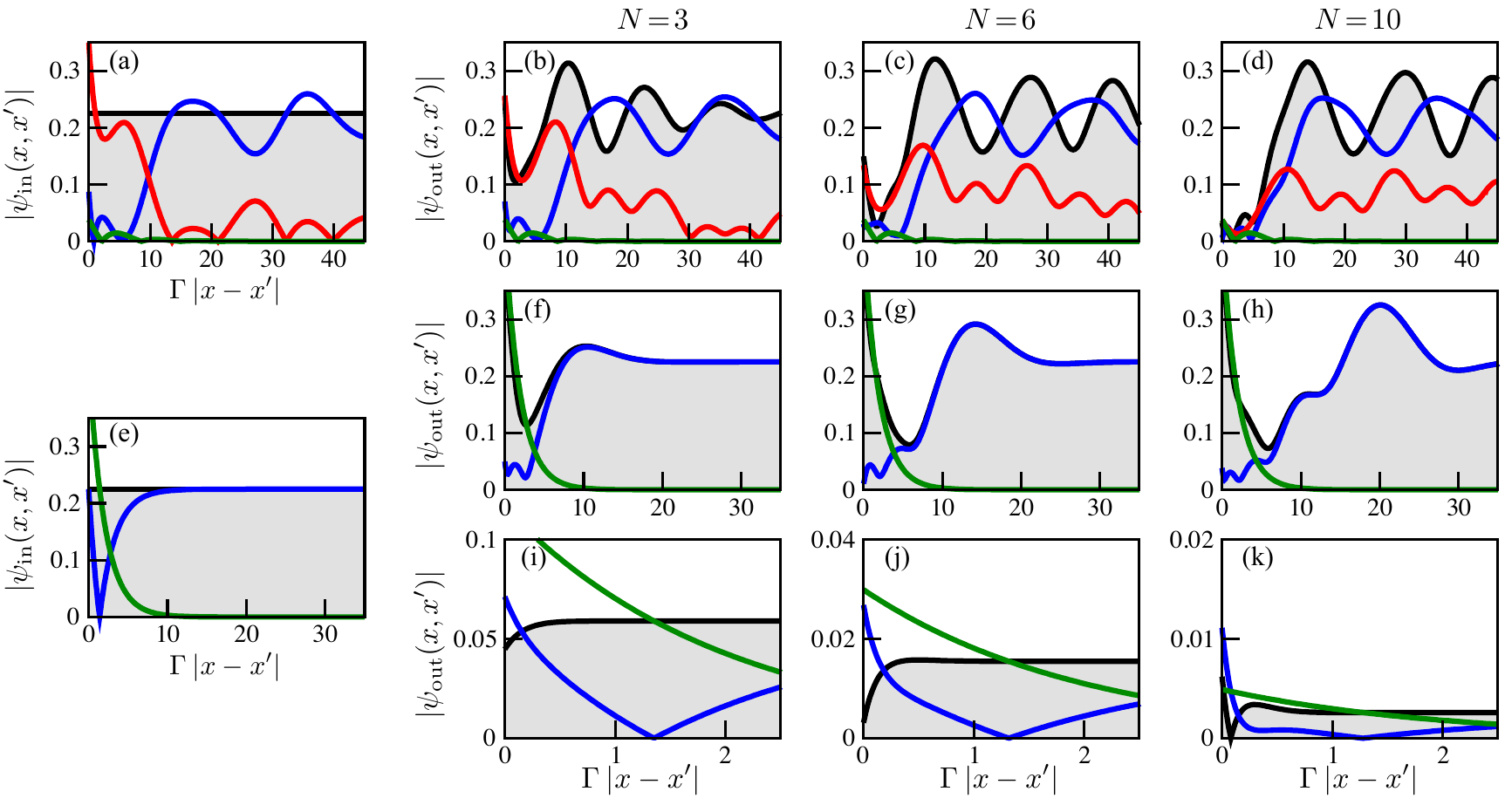}
	\caption{Decomposition of incident state [(a) and (e)] and transmitted photon states [(b)--(d) and (f)--(h)] into the three types of eigenstates of the $S$-matrix: ($i$) superposition of plane waves (blue lines), ($ii$) superposition of plane waves and an exponentially localized contribution (red lines), and ($iii$) the two-photon bound state (green), for a chain of three-level systems (a)--(d) and a chain of two-level systems (e)--(h). The total wave function is indicated by the black line and grey shaded area. Note that the type ($ii$) of eigenstates does not exist for two-level emitters. In panels (i)--(k) we show the output for a dissipative ($\gamma = 10\Gamma$) two-level chain where the input field is on resonance. In panels (a)--(h) we use single-photon detuning $\Delta = \Gamma/4$, and we use $\Omega = \Gamma/2$ and $\bar{\Delta} = -\Gamma/4$ for the three-level system [(a)--(d)]. Panel (e) shows the decomposition of the input for all panels (f)--(k). \label{fig:contributions}}
\end{figure*}

Without photon loss ($\gamma = 0$), we here determine the complete set of two-photon eigenstates \cite{sf07,sf07-prl} of the underlying scattering matrix ($S$-matrix), which connects the outgoing scattered state $|{\rm out}\rangle=\hat{S}|{\rm in}\rangle$ to the incident state $|{\rm in}\rangle$ of the two photons before their interaction with the emitter. We find three qualitatively distinct classes of eigenstates $\ket{E,\nu}$ that depend on the two-photon energy $E$ and the quantum number $\nu$, which is a measure of the relative momentum of the two photons \cite{SM}. In addition to typical scattering states that can be written as superpositions of plane wave solutions, the light-matter coupling leads to a two-photon bound state in which the distance between the outgoing photons is localized exponentially. While these types of states are also found for two-level emitters \cite{sf07,sf07-prl}, the optical coupling to a third level leads to another type of continuum states, which consist of an exponentially localized contribution and fully delocalized plane waves [Fig. \ref{fig:system}(c)]. As we shall see below, these states play an important role for the characteristic transmission properties of the chain. 

Having obtained the complete set of eigenstates from $\hat{S}|E,\nu\rangle=\lambda(E,\nu)|E,\nu\rangle$, along with their eigenvalues $\lambda(E,\nu) = e^{i\varphi_{E,\nu}}$, we can use the spectral decomposition of the $S$-matrix to determine the two-photon output 
\begin{equation}
	\ket{\textup{out}} = \int dE \mathrlap{\sum}\int\limits_{\nu} \lambda(E,\nu)^N \ket{E,\nu} \langle E,\nu \vert \textup{in}\rangle, \label{eq:decompose}
\end{equation}
at the end of the chain, by an $N$-fold application of the single-particle $S$-matrix \cite{Mahmoodian_2018}. Here, $\mathrlap{\sum}\int$ includes the discrete photon bound state and the integral over the quantum number $\nu$ of the continuum states. 

Up to a normalization constant, the obtained two-photon wave function $\langle x,x^\prime|{\rm out}\rangle$ at the end of the chain ($x,x^\prime>x_N$) yields the two-photon correlation function $g^{(2)}(r)\propto |\langle x,x^\prime|{\rm out}\rangle|^2$ which only depends on the relative distance $r=x-x^\prime$ between the photons but not their center of mass coordinate. An example is shown in Fig. \ref{fig:system}(a) for a moderately sized chain of $N=10$ emitters. Evidently, the incident product state is converted into a highly correlated two-photon output with virtually complete antibunching, indicating that the photons leave the chain with a large delay of more than $5/\Gamma$. The generation of such strong antibunching requires the action of several emitters, and we observe a clear transition from photon bunching ($g^{(2)}(0) > 1$) to antibunching ($g^{(2)}(0) < 1$) as we increase the length of the chain. As shown in Fig. \ref{fig:system}(b), one can identify a critical number $N$ beyond which the transmitted light remains strongly antibunched regardless of $N$. The underlying process is therefore robust against number fluctuations, such that precise control of the chain length is not essential to generate single photons in our setup. 

We can gain a better understanding of the underlying mechanism by considering the decomposition of the uncorrelated two-photon input into the three different types of eigenstates, as shown in Fig. \ref{fig:contributions}(a). As the photons scatter off an increasing number of emitters, the $S$-matrix eigenstates with continuous quantum numbers $\nu$ pick up different phases and start to dephase [cf. Figs. \ref{fig:contributions}(b)--(d)], as described by the growing phase of $\lambda(E,\nu)^N ={\rm e}^{iN\varphi_{E,\nu}}$ in Eq. (\ref{eq:decompose}). Upon increasing $N$, the value of $g^{(2)}(0)$ will therefore eventually be dominated by the initial bound state contribution, as it remains unaffected by dephasing. Importantly, the coupling to the third meta-stable state leads to a very small bound state component of the initial state, whose contribution is replaced by the new scattering eigenstates that also feature exponentially decaying behavior at small distances. This makes strong antibunching possible and differs from the case of two-level systems, where the bound state contributes significantly and therefore causes bunching of the transmitted photons [cf. Figs. \ref{fig:contributions}(e)--(h)]. Yet, antibunching can still be achieved with two-level emitters \cite{Mahmoodian_2018,Prasad_2019} in a finite interval of particle numbers by adding dissipation at the cost of an overall reduced transmission [cf. Figs. \ref{fig:contributions}(i)--(k)].

\begin{figure}
		\includegraphics[width=0.92\columnwidth]{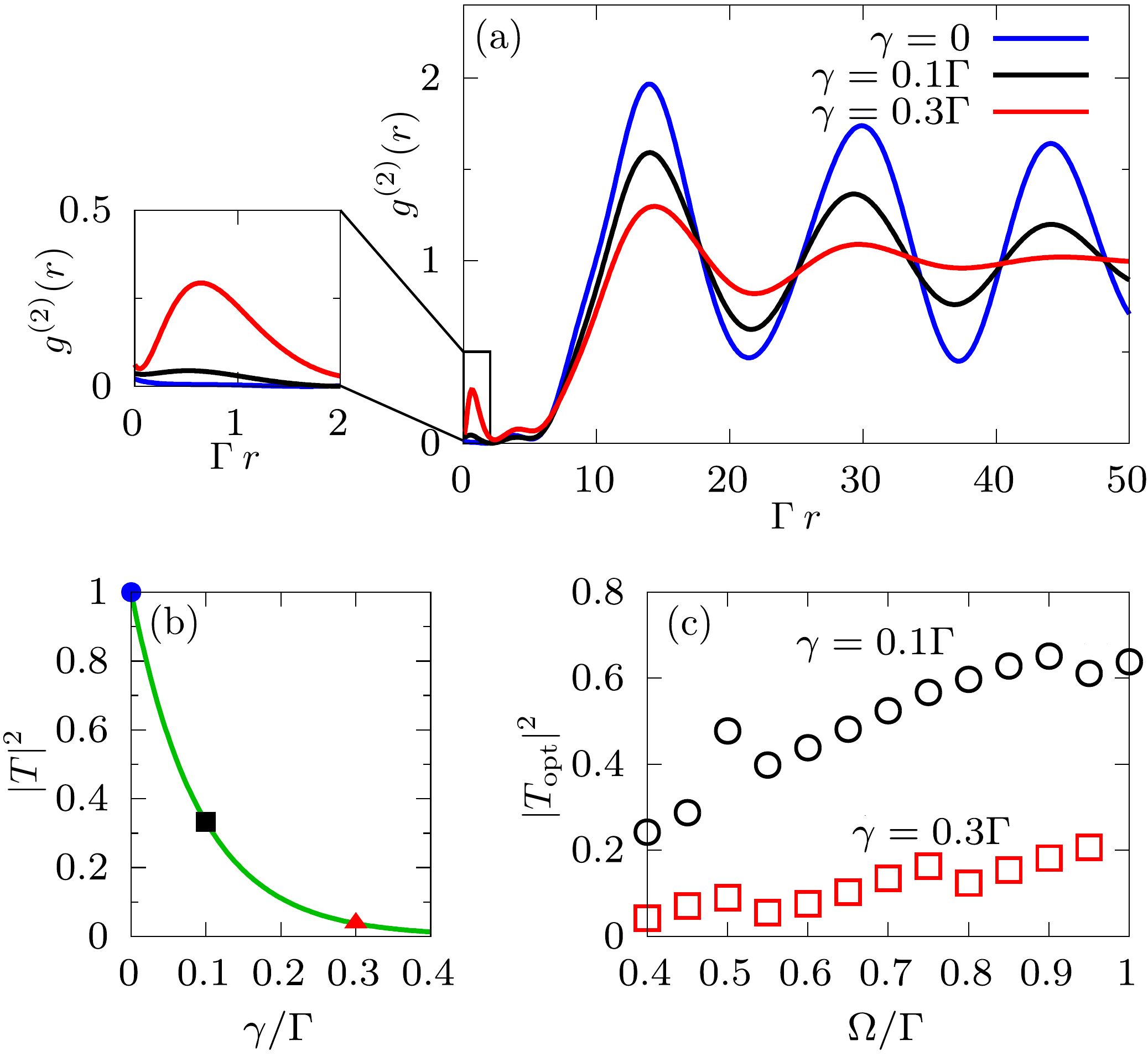}
	\caption{(a) The two-photon correlation function $g^{(2)}(r)$ for $\Omega = \Gamma/2$, $\Delta = \Gamma/4$, $\bar{\Delta} = -\Gamma/4$, $N = 10$, and different decay rates $\gamma$ into other modes than the guided. (b) Two photon transmission through the chain of 10 emitters with the parameters of panel (a) as a function of $\gamma$. The points mark the transmission for the values of $\gamma$ used in panel (a). (c) Optimal two-photon transmission, $|T_{\rm opt}|^2$, as a function of $\Omega$ for $\Delta = 0.4\Gamma$ and $\bar{\Delta} = -0.2\Gamma$, obtained for the minimal number of emitters required to achieve $g^{(2)}(0) \leq 0.1$. \label{fig:dissipative}}
\end{figure}

While photon losses into non-guided modes can be suppressed substantially, e.g. using photonic crystal waveguides \cite{chiral_crystal,Arcari_2014}, dissipation inevitably occurs in experiments and is accounted for by the decay rate $\gamma$ in Eq. (\ref{eq:H}). As finite losses render the corresponding $S$-matrix non-unitary and jeopardize orthogonality of its eigenstates, we employ here another approach to describe the nonlinear photon propagation for $\gamma>0$. Hereby, we treat the chain as a cascaded quantum system and calculate the wave function of two photons due to their interaction with a given emitter \cite{zgb10,zgb12} at position $x_j$, using the two-photon outgoing state of the prior emitter at position $x_{j-1}$ as an incoming boundary conditions. This yields a recursion relation for the successive photon output from each emitter \cite{SM}, which can be evaluated numerically for moderate values of $N$ to obtain the correlated two-photon transmission of the entire chain. Fig. \ref{fig:dissipative}(a) shows the two-photon correlations for different values of $\gamma$. The results demonstrate that the generated antibunching indeed survives a finite amount of dissipation until the loss rate eventually becomes too large, and one eventually observes photon bunching.

A more drastic consequence of dissipation is therefore the reduction of the overall two-photon transmission 
\begin{equation}
	|T_N|^2 = \left|\frac{\Omega^2 - \Delta\bar{\Delta} + i(\gamma-\Gamma)\bar{\Delta}/2}{\Omega^2 - \Delta\bar{\Delta} + i(\gamma+\Gamma)\bar{\Delta}/2}\right|^{2N}, \label{eq:TNsq}
\end{equation}
which can be obtained from the independent transmission \cite{ws10} for each photon. Even though the simultaneous transmission of two photons remains highly suppressed with respect to the overall transmission [Fig. \ref{fig:dissipative}(a)], its significant drop [Fig. \ref{fig:dissipative}(b)] limits the efficiency of single-photon generation. 

It turns out, however, that the additional control field coupling $\Omega$ offers an effective means to control photon dissipation in the system. Hereby, we can exploit the emergence of electromagnetically induced transparency (EIT) \cite{Fleischhauer2005} when the system is tuned to two-photon resonance ($\bar{\Delta} = 0$), suppressing dissipation from the intermediate level $|e\rangle$. While the generation of strong photon interactions requires to avoid perfect EIT conditions \cite{rb14,fb16}, it is still possible to decrease dissipation effects by operating close to two-photon resonance. Since the EIT window broadens with the control field Rabi frequency $\Omega$ \cite{ws10,roy11,zgb12,rb14}, increasing $\Omega$ is expected to decrease photon losses. Working closer to perfect EIT conditions, on the other hand, tends to weaken the optical nonlinearity such that longer chains are required to reach a given level of antibunching, which in turn increases the overall losses. In order to explore this interesting competition, we have determined the minimum chain length $N_{\rm opt}$ required to obtain $g^{(2)}(0) \leq 0.1$ for different values of $\Omega$ \cite{SM}. Using this optimal chain length in Eq. (\ref{eq:TNsq}) yields the achievable transmission of antibunched photons, as shown in Fig. \ref{fig:dissipative}(c) and indicates that photon losses can indeed be greatly suppressed by exploiting EIT in the present system.

Experimentally, efficient coupling between quantum emitters and nano-scale waveguides can be achieved by interfacing atoms \cite{Hood2016} or quantum dots \cite{chiral_crystal,Arcari_2014} with photonic crystal waveguides, or with atoms near optical nanofibers \cite{Petersen_2014,Mitsch_2014}. In particular, the coupling of emitters with transition frequencies close to the band gap of photonic crystal structures can in principle yield arbitrarily large coupling efficiencies $\beta=\Gamma/(\gamma+\Gamma)$ \cite{John90}, and has been used to realize strong photon coupling with $\beta\sim 0.6$ for atomic interfaces \cite{Hood2016} and $\beta\sim 0.98$ in quantum dot experiments \cite{Arcari_2014}. Moreover, EIT has been demonstrated with atoms in numerous experiments \cite{Boller1991,Fleischhauer2005}, while the considered three-level $\Lambda$-scheme can also be implemented in solid state settings, such as charged quantum dots \cite{Warburton_2013}.

In conclusion, we have studied two photon transport through a chiral waveguide coupled to a chain of three-level emitters, and found that the collective coupling of multiple emitters can generate strongly antibunched light when the length of the chain exceeds a critical size. The underlying mechanism does not rely on nonlinear dissipation or strong atomic interactions, and neither requires regularly spaced emitters or sub-wavelength distances. Since it is based entirely on interference, one can exploit EIT to generate strong photon correlations under greatly suppressed losses. This proves useful for the efficient generation of single photons, i.e. the conversion of a classical input field into a stream of antibunched light with high transmission. Hereby, the long-range correlations and persistent oscillations of $g^{(2)}$ [cf. Fig.\ref{fig:system}(a)] found for two interacting photons, suggest interesting perspectives for explorations of strongly correlated many-body states of light, and motivates future experimental and theoretical work to understand the dynamics of multi-photon quantum states in the proposed setting.

We thank Sebastian Hofferberth and Klaus Mølmer for helpful comments and fruitful discussions. 
This work was supported by the Carlsberg Foundation through the 'Semper Ardens' Research Project QCooL, by the DFG through the SPP1929, by the European Commission through the H2020-FETOPEN project ErBeStA (No. 800942), and by the DNRF through a Niels Bohr Professorship to TP.
\onecolumngrid
\appendix
\section{Equations of motion and boundary conditions}
Here we present a more detailed description of the derivation of the amplitude equations describing two-photon scattering off a single atom. The Hamiltonian for a single emitter is given by
\begin{equation}
\begin{split}
		\hat{h} = \int {\rm d}x (-i)\hat{c}^{\dagger}(x)\partial_{x} \hat{c}(x) + \left[\omega_{2} - i\frac{\gamma}{2}\right]\hat{\sigma}_{ee} + (\omega_3 + \bar{\omega})\hat{\sigma}_{bb} + \left[\int dx V\delta(x) \hat{c}(x)\hat{\sigma}_{ea} + \Omega \hat{\sigma}_{be} + \textup{h.c.}\right], \label{eq:Hsingle}
	\end{split} 
\end{equation}
where $V = \sqrt{\Gamma}$ is the atom-waveguide coupling \cite{fks10}. As described in the main text, we use the following ansatz 
\begin{equation}
		\ket{\phi} = \bigg(\int {\rm d}x{\rm d}x' \psi(x,x')\frac{1}{\sqrt{2}}\hat{c}^{\dagger}(x)\hat{c}^{\dagger}(x') + \int {\rm d}x\left[e(x)\hat{\sigma}_{ea} + b(x)\hat{\sigma}_{ba}\right]\hat{c}^{\dagger}(x) \bigg) \vert 0,a\rangle. \label{eq:scattering-state}
\end{equation}
to derive the equations of motion, where $\vert 0, a\rangle$ is the state with zero photons and the emitter in the ground state. Requiring that $\vert \phi\rangle$ is an eigenstate of the Hamiltonian \eqref{eq:Hsingle} with energy $E$ one obtains the following equations (see Ref. \cite{sf07} for a detailed description for two-level systems)
\begin{subequations}
	\label{eq:equations-of-motion}
	\begin{align}
		\left[-i\partial_{x} - i\partial_{x'} - E\right]\psi(x,x') + \frac{V}{\sqrt{2}}\left[\delta(x)e(x') + \delta(x')e(x)\right] &= 0, \label{eq:equations-of-motion1} \\
		\left[-i\partial_{x} + \omega_2 - i\frac{\gamma}{2} - E\right]e(x) + \frac{V}{\sqrt{2}}\left[\psi(0,x) + \psi(x,0)\right] + \Omega b(x) &= 0, \label{eq:equations-of-motion2} \\
		\left[-i\partial_{x} + \omega_3 + \bar{\omega} - E\right]b(x) + \Omega e(x) &= 0, \label{eq:equations-of-motion3}
	\end{align}
\end{subequations}
for the different amplitudes, where
\begin{equation}
	\psi(x,0) = \psi(0,x) = \frac{1}{2}\left[\psi(0^-,x) + \psi(0^+,x)\right]
\end{equation}
From Eqs. \eqref{eq:equations-of-motion} we obtain the boundary conditions for $x < x'$
\begin{subequations}
	\label{eq:boundary}
	\begin{align}
		-i\left[\psi(x,0^+) - \psi(x,0^-)\right] + \frac{V}{\sqrt{2}}e(x) &= 0 \quad (x < 0), \label{eq:boundary1} \\
		-i\left[\psi(0^+,x') - \psi(0^-,x')\right] + \frac{V}{\sqrt{2}}e(x') &= 0 \quad (x' > 0), \label{eq:boundary2} \\
		e(0^+) &= e(0^-), \label{eq:boundary3} \\
		b(0^+) &= b(0^-). \label{eq:boundary4}
	\end{align}
\end{subequations}
Apart from the boundary conditions in Eqs. \eqref{eq:boundary} we require that $\psi(x,x')$, $e(x)$, and $b(x)$ are continuous functions when $x,x'\neq 0$ \cite{sf07}.

Now we note that when $x,x'\neq 0$, it follows from Eq. \eqref{eq:equations-of-motion1} that $\psi(x,x') \propto e^{iEr_c}$, where $r_c \equiv (x+x')/2$ is the center of mass coordinate. Therefore $\psi(x,x')$ must have the general form
\begin{equation}
	\psi(x,x') = e^{iEr_c} F(r), \label{eq:general-form}
\end{equation}
where $r \equiv x - x'$ is the relative coordinate, and $F(r)$ is continuous when $x,x'\neq 0$. The main features of $\psi(x,x')$ are hidden in $F(r)$ since $e^{iEr_c}$ is only a phase that varies with the center of mass position, but not the distance between the photons. We see that $F(r)$ must have the form
\begin{equation}
F(r) = \begin{cases}
	F_{\textup{in}}(r), & x < x' < 0, \\
	F_{\textup{0}}(r), & x < 0 < x', \\ 
	F_{\textup{out}}(r) & x' > x > 0,
\end{cases} \label{eq:F-form}
\end{equation}
where the part for $x > x'$ ($r > 0$) is given by bosonic symmetry.

We now derive a set of equations for $F(r)$ when $x < x'$. First we substitute Eqs. \eqref{eq:general-form} and \eqref{eq:F-form} into Eqs. \eqref{eq:boundary1} and \eqref{eq:boundary2} to obtain
\begin{subequations}
\label{eq:e2}
	\begin{align}
		e(x) &= \frac{\sqrt{2}i}{V} \left[F_{\textup{0}}(x) - F_{\textup{in}}(x)\right] e^{iEx/2} \quad (x < 0), \label{eq:e2_x1}\\
		e(x') &= \frac{\sqrt{2}i}{V} \left[F_{\textup{out}}(-x') - F_{\textup{0}}(-x')\right] e^{iEx'/2} \quad (x' > 0).
	\end{align}
\end{subequations}
Using Eqs. \eqref{eq:general-form}, \eqref{eq:F-form}, and \eqref{eq:e2_x1} in Eqs. \eqref{eq:equations-of-motion2} and \eqref{eq:equations-of-motion3} yields for $x < 0$
\begin{subequations}
	\label{eq:F-e_3-system}
	\begin{align}
		\left[\partial_{x} + i\Delta + \frac{\gamma+\Gamma}{2}\right]F_{\textup{0}}(x) - \left[\partial_{x} + i\Delta + \frac{\gamma-\Gamma}{2}\right] F_{\textup{in}}(x) + \frac{V\Omega}{\sqrt{2}}b(x)e^{-iEx/2} &= 0, \\
		\left[\partial_{x} + i\bar{\Delta}\right] \left[b(x)e^{-iEx/2}\right] - \frac{\sqrt{2}\Omega}{V}\left[F_{\textup{0}}(x) - F_{\textup{in}}(x)\right] &= 0,
	\end{align}
\end{subequations} 
where $\Delta = \omega_e - \omega$ is the single-photon detuning when two incident photons have the same energies $\omega = E/2$, and $\bar{\Delta} = \omega_b - (\omega - \bar{\omega})$ is the two-photon detuning. Finally, $b(x)$ can be eliminated in Eqs. \eqref{eq:F-e_3-system} to obtain a single second order differential equation
\begin{equation}
	\left[\partial_r^2 + \eta\partial_r + \alpha\right] F_{\textup{0}}(r) = \left[\partial_r^2 + \eta\partial_r + \alpha\right]F_{\textup{in}}(r) + \left[\mu - \Gamma\partial_r\right]F_{\textup{in}}(r) \quad (r < 0), \label{eq:F1}
\end{equation}
where 
\begin{align}
	\alpha &\equiv \Omega^2 + i\bar{\Delta}\left(i\Delta + \frac{\gamma + \Gamma}{2}\right), \label{eq:alpha} \\
	\eta &\equiv i(\Delta+\bar{\Delta}) + \frac{\gamma+\Gamma}{2}, \\
	\mu &\equiv -i\Gamma\bar{\Delta}. \label{eq:mu}
\end{align}
Similarly, it can be shown that
\begin{equation}
	\left[\partial_r^2 - \eta\partial_r + \alpha\right] F_{\textup{out}}(r) = \left[\partial_r^2 - \eta\partial_r + \alpha\right]F_{\textup{0}}(r) + \left[\mu + \Gamma\partial_r\right]F_{\textup{0}}(r) \quad (r < 0). \label{eq:F2}
\end{equation}
Using Eqs. \eqref{eq:equations-of-motion2} and \eqref{eq:e2} one finds that the boundary conditions \eqref{eq:boundary3} and \eqref{eq:boundary4} yield
\begin{subequations}
	\label{eq:boundary-F}
	\begin{align}
		F_{\textup{out}}(0) &= 2F_{\textup{0}}(0)-F_{\textup{in}}(0), \\
		\partial_r F_{\textup{out}}(r)\vert_{r=0} &= \partial_r F_{\textup{in}}(r)\vert_{r=0} + \Gamma\left[F_{\textup{0}}(0) - F_{\textup{in}}(0)\right].
	\end{align}
\end{subequations}
Eqs. \eqref{eq:F1}, \eqref{eq:F2}, and \eqref{eq:boundary-F} constitutes the set of equations we will use to solve the scattering problem in the non-dissipative and dissipative case in the following two sections.

\section{Description of unitary photon propagation}
In this section we show how we find the outgoing two-photon wave function of the chain using the eigenstates of the single-atom scattering matrix ($S$-matrix). As noted in the main text, the two-photon correlation function can be found directly from this outgoing wave function. In short, we find the eigenstates of the single-atom $S$-matrix and decompose our incident state into these eigenstates \cite{sf07,sf07-prl}. With this decomposition at hand, the outgoing state after all $N$ atoms is found by multiplying the eigenstates by their eigenvalues to the $N$th power in the decomposition \cite{Mahmoodian_2018}. Throughout this section we assume that there is no decay to other modes than the guided one ($\gamma =0$). 

To find the eigenstates of the $S$-matrix we start by eliminating $F_{\textup{0}}(r)$ in Eqs. \eqref{eq:F1}, and \eqref{eq:F2}, and thereby obtain the fourth order differential equation
\begin{equation}
	\left[\partial_r^4 - (\eta^2-2\alpha)\partial_r^2 + \alpha^2\right]F_{\textup{out}}(r)
	= \left[\partial_r^4 - \left[(\Gamma - \eta)^2-2(\alpha + \mu)\right]\partial_r^2 + (\alpha+\mu)^2\right]F_{\textup{in}}(r). \label{eq:diff-Fout-Fin}
\end{equation}
To find eigenstates of the $S$-matrix we impose the condition $F_{\textup{out}}(r) = \lambda F_{\textup{in}}(r)$, where $\lambda$ is the eigenvalue of the $S$-matrix, in Eq. \eqref{eq:diff-Fout-Fin}
\begin{equation}
	\lambda\left[\partial_r^4 - (\eta^2-2\alpha)\partial_r^2 + \alpha^2\right]F_{\textup{in}}(r)
	= \left[\partial_r^4 - \left[(\Gamma - \eta)^2-2(\alpha + \mu)\right]\partial_r^2 + (\alpha+\mu)^2\right]F_{\textup{in}}(r). \label{eq:diff-Fin}
\end{equation}
This equation is solved by the form
\begin{equation}
	F_{\textup{in}}(r) = A e^{-i\nu x} + B e^{i\nu r} + C e^{-i\tilde{\nu} r} + De^{i\tilde{\nu} r}, \label{eq:Fin-full}
\end{equation}
where $A$, $B$, $C$, and $D$ are constants to be found. Note that for a two-level atom the form would be \cite{sf07,sf07-prl}
\begin{equation}
	F_{\textup{in}}^{(\textup{two-level})}(r) = A' e^{-i\nu r} + B' e^{i\nu r},
\end{equation}
where $\nu$ is half the momentum difference of the two photons. Our interpretation of Eq. \eqref{eq:Fin-full} is therefore that the eigenstates for the three-level system mix two different values ($\nu$ and $\tilde{\nu}$) of relative momenta, which means that states with momenta $k,p$ hybridize with another set $\tilde{k},\tilde{p}$, where the total energy $k + p = \tilde{k} + \tilde{p}$ is the same for both [by Eq. \eqref{eq:general-form}].

Substituting Eq. \eqref{eq:Fin-full} into Eq. \eqref{eq:diff-Fin} shows that the eigenvalue of the $S$-matrix is given by
\begin{equation}
	\lambda(E,\nu) = \frac{\nu^4 + \left[(\Gamma - \eta)^2-2(\alpha + \mu)\right]\nu^2 + (\alpha+\mu)^2 }{\nu^4 + (\eta^2-2\alpha)\nu^2 + \alpha^2} = \frac{\tilde{\nu}^4 + \left[(\Gamma - \eta)^2-2(\alpha + \mu)\right]\tilde{\nu}^2 + (\alpha+\mu)^2 }{\tilde{\nu}^4 + (\eta^2-2\alpha)\tilde{\nu}^2 + \alpha^2} = \lambda(E,\tilde{\nu}). \label{eq:a}
\end{equation}
This means that for a given $\nu$ the value of $\tilde{\nu}$ is fixed, and we can find it by solving the equation above with the constraint $\tilde{\nu} \neq \pm \nu$. This gives
\begin{equation}
	\tilde{\nu}^2 = \frac{\mu(2\alpha+\mu)\nu^2 + \alpha^2\left[\Gamma(2\eta-\Gamma)+2\mu\right] + (\eta^2 - 2\alpha)\mu(2\alpha+\mu)}{\left[\Gamma(2\eta-\Gamma)+2\mu\right]\nu^2 - \mu(2\alpha+\mu)}. \label{eq:nutilde}
\end{equation}
In the absence of dissipation ($\gamma = 0$) it follows from Eq. \eqref{eq:nutilde} that $\tilde{\nu}^2$ is real when $\nu^2$ is real. Note, moreover, that Eq. \eqref{eq:nutilde} leaves an ambiguity in the sign of $\tilde{\nu}$. To get rid of this ambiguity we define the phase of $\tilde{\nu}$ to be in the interval between $0$ and $\pi$, i.e. $0 \leq \operatorname{arg}[\tilde{\nu}] < \pi$.

It is worth noting that when the system is non-dissipative, we must have $|\lambda(E,\nu)|^2 = 1$ (perfect transmission), which implies $\lambda(E,\nu)^* = \lambda(E,\nu)^{-1}$. Using that $\alpha^* = \alpha + \mu$ and $\eta^* = \Gamma - \eta$ when $\gamma = 0$ together with Eq. \eqref{eq:a}, it follows that
\begin{equation}
	{\nu^*}^2 = \nu^2 \quad \textup{or} \quad {\nu^*}^2 = \tilde{\nu}^2.
\end{equation}
In particular, any real value of $\nu$ gives an allowed value of the eigenvalue $\lambda(E,\nu)$.

The actual eigenstates of the $S$-matrix have [by Eq. \eqref{eq:general-form}] real space representations (or wave functions)
\begin{equation}
	\Psi_{E,\nu}(x,x') = \langle x,x'\ket{E',\nu} = e^{iE r_c}F_{E,\nu}(r), \label{eq:eig-realspace}
\end{equation}
where $F_{E,\nu}(r)$ is given by Eq. \eqref{eq:Fin-full}. These states are shown exemplarily in Fig. 1(c) of the main text. \\

To find the constants $A$, $B$, $C$, and $D$ in the eigenstates given by Eq. \eqref{eq:Fin-full}, we apply the boundary conditions \eqref{eq:boundary-F}. For this purpose we need $F_{\textup{0}}(r)$, which we find by solving Eq. \eqref{eq:F1} to obtain
\begin{equation}
	\begin{split}
		F_{\textup{0}}(r) &= \frac{\nu^2 - i(\Gamma - \eta)\nu - (\alpha + \mu)}{\nu^2 + i\eta\nu - \alpha}Ae^{-i\nu r} + \frac{\nu^2 + i(\Gamma - \eta)\nu - (\alpha + \mu)}{\nu^2 - i\eta\nu - \alpha}Be^{i\nu r} \\
			&\quad+ \frac{\tilde{\nu}^2 - i(\Gamma - \eta)\tilde{\nu} - (\alpha + \mu)}{\tilde{\nu}^2 + i\eta\tilde{\nu} - \alpha}Ce^{-i\tilde{\nu} r} + \frac{\tilde{\nu}^2 + i(\Gamma - \eta)\tilde{\nu} - (\alpha + \mu)}{\tilde{\nu}^2 - i\eta\tilde{\nu} - \alpha}De^{i\tilde{\nu} r}.
	\end{split}
\end{equation}
Since there are only two boundary conditions and the constraint of normalization, there is a free parameter in Eq. \eqref{eq:Fin-full}. To still find an expression we initially set $D = 0$. In this case we find that the eigenstate is
\begin{equation}
	F_{E,\nu}^{(D=0)}(r) = \frac{1}{\xi(E,\nu)}\left[u_E(\tilde{\nu},-\nu)h_E(\nu)e^{-i\nu r} + u_E(\nu,\tilde{\nu})h_E(\nu)e^{i\nu r} - u_E(\nu,-\nu)h_E(\tilde{\nu})e^{-i\tilde{\nu} r}\right], \label{eq:F_D=0}
\end{equation}
where
\begin{align}
	\begin{split}
			u_E(\nu,\nu') &= \left[\Gamma(\Gamma\alpha+\eta\mu) + \mu^2\right](\nu-\nu')\big[2(2\eta-\Gamma)\nu\nu'(\nu+\nu') \\
				&\quad + 2i\mu(\nu^2 + \nu\nu' + {\nu'}^2) - i\Gamma(2\eta-\Gamma)\nu\nu' + \Gamma\mu(\nu+\nu') - i\mu(2\alpha+\mu)\big], 
	\end{split}\\
	h_E(\nu) &= \nu^4 + (\eta^2 - 2\alpha)\nu^2 + \alpha^2,
\end{align}
and 
\begin{equation}
	\begin{split}
		\xi(E,\nu) &= 2\pi\left|h_E(\nu)\right| \Big(|u_E(\tilde{\nu},-\nu)|^2 + |u_E(\nu,\tilde{\nu})|^2 + 4\left[\Gamma(\Gamma\alpha + \eta\mu) + \mu^2\right]^2\textup{Re}[\tilde{\nu}] \\
			&\quad \times \left|\nu\left[\left[\Gamma(2\eta-\Gamma)+2\mu\right]\tilde{\nu}^2 - \mu(2\alpha+\mu)\right]\left[\left[\Gamma(2\eta-\Gamma)+2\mu\right]\nu^2 - \mu(2\alpha+\mu)\right]\right|\Big)^{1/2}
	\end{split}
\end{equation}
is a normalization factor. By setting $C = 0$ instead, we get
\begin{equation}
	F_{E,\nu}^{(C=0)}(r) = \frac{1}{\xi(E,\nu)}\left[u_E(-\tilde{\nu},-\nu)h_E(\nu)e^{-i\nu r} + u_E(\nu,-\tilde{\nu})h_E(\nu)e^{i\nu r} - u_E(\nu,-\nu)h_E(\tilde{\nu})e^{i\tilde{\nu} r}\right]. \label{eq:F_C=0}
\end{equation}

In order to find all eigenstates, we go through the different scenarios for $\nu$ and $\tilde{\nu}$. First, we consider the case where $\nu$ is real. If $\tilde{\nu}$ is not real, we have $\operatorname{Im}[\tilde{\nu}] > 0$ by our choice of phase ($0 \leq \operatorname{arg}[\tilde{\nu}] < \pi$), and because $\tilde{\nu}^2$ is real, we must have $\operatorname{Re}[\tilde{\nu}] = 0$ in this case. Then Eq. \eqref{eq:F_D=0} gives the only physical eigenstate since Eq. \eqref{eq:F_C=0} diverges when $r \to -\infty$. These states are the continuum eigenstates with wave functions that are superpositions of plane waves and a localized term, mentioned in the main text, that emerge due to the coupling to the third level $| b \rangle$.

Conversely, if both $\nu$ and $\tilde{\nu}$ are real, both Eq. \eqref{eq:F_D=0} and Eq. \eqref{eq:F_C=0} describe two different valid states, which are degenerate both in energy and in the eigenvalue of the $S$-matrix. Therefore they need not be orthogonal. Another point of view is to note that when both $\nu$ and $\tilde{\nu}$ are real, these quantum numbers can be interchanged in Eq. \eqref{eq:F_D=0}, which gives rise to a new eigenstate $F_{E,\tilde{\nu}}^{(D=0)}(r)$ with the same eigenvalue as $F_{E,\nu}^{(D=0)}(r)$, and therefore these states need not be orthogonal. Instead we can use
\begin{equation}
	F_{E,\nu}^{(\textup{I})}(r) = \frac{1-\zeta}{2}F_{E,\nu}^{(D=0)}(r) + \frac{1+\zeta}{2}\left[\theta(\nu^2 - \tilde{\nu}^2)F_{E,\nu}^{(D=0)}(r) + \theta(\tilde{\nu}^2 - \nu^2)F_{E,\nu}^{(C=0)}(r)\right], \label{eq:F_Enu_I}
\end{equation}
or
\begin{equation}
	F_{E,\nu}^{(\textup{II})}(r) = \frac{1-\zeta}{2}F_{E,\nu}^{(C=0)}(r) + \frac{1+\zeta}{2}\left[\theta(\nu^2 - \tilde{\nu}^2)F_{E,\nu}^{(C=0)}(r) + \theta(\tilde{\nu}^2 - \nu^2)F_{E,\nu}^{(D=0)}(r)\right], \label{eq:F_Enu_II}
\end{equation}
when $\tilde{\nu}$ is real, and where
\begin{equation}
	\zeta = \textup{sgn}\left[\frac{\left[\Gamma(2\eta-\Gamma)+2\mu\right]\tilde{\nu}^2 - \mu(2\alpha+\mu)}{\left[\Gamma(2\eta-\Gamma)+2\mu\right]\nu^2 - \mu(2\alpha+\mu)}\right],
\end{equation}
where $\textup{sgn}$ is the sign function. The states represented by $F_{E,\nu}^{(\textup{I})}(r)$ and $F_{E,\tilde{\nu}}^{(\textup{I})}(r)$ are orthogonal, and similarly the states represented by $F_{E,\nu}^{(\textup{II})}(r)$ and $F_{E,\tilde{\nu}}^{(\textup{II})}(r)$ are orthogonal, so these states thereby ensure orthogonality. The states \eqref{eq:F_Enu_I} and \eqref{eq:F_Enu_II} are the continuum states that consist of only plane waves, mentioned in the main text. 

Note that the Eq. \eqref{eq:F_Enu_I} can also be used as the eigenstate for complex values of $\tilde{\nu}$, since we automatically have $\tilde{\nu}^2 < \nu^2$ in that case. The two continuum states shown in Fig. 1(c) of the main text have the form of Eq. \eqref{eq:F_Enu_I} with different real values of $\nu$ such that $\tilde{\nu}$ is real for the eigenstates consisting only of plain waves, and $\tilde{\nu}$ is complex for the eigenstates consisting of plane waves and a localized term.

All the states found so far are continuum states, i.e. they exist for a continuous set of quantum numbers $E$ and $\nu$, and their real space representations do not vanish for $r \to \pm\infty$. It is, however, known that for a two-level system, the continuum states do not constitute a complete set, but a bound state, with the property that the real space representation vanishes for $r \to \pm\infty$, is needed at every energy \cite{sf07,sf07-prl}. To search for such an eigenstate in the present setting, we use that such an eigenstate must still have the form \eqref{eq:Fin-full}, but with complex $\nu$ and $\tilde{\nu}$. With $0 < \arg[\nu],\arg[\tilde{\nu}] < \pi$ we find the bound state numerically by requiring that $B = D = 0$ (so the state can be normalized) in \eqref{eq:Fin-full}. This state is a bound state with a value of $\nu$, which we denote by $\nu_b$. The eigenstate is in this case given by
\begin{equation}
	F_{E,\nu_b}(r) = \frac{1}{\xi_b(E,\nu_b)}\left[u_E(\tilde{\nu}_b,-\nu_b)h_E(\nu_b)e^{-i\nu_b r} - u_E(\nu_b,-\nu_b)h_E(\tilde{\nu}_b)e^{-i\tilde{\nu}_b r}\right], \label{eq:F_bound}
\end{equation} 
where 
\begin{equation}
	\xi_b(E,\nu_b) = 2\sqrt{\pi}\sqrt{\frac{\left|u_E(\tilde{\nu}_b,-\nu_b)h_E(\nu_b)\right|^2}{2\operatorname{Im}[\nu_b]} + \frac{\left|u_E(\nu_b,-\nu_b)h_E(\tilde{\nu}_b)\right|^2}{2\operatorname{Im}[\tilde{\nu}_b]} - 2\operatorname{Im}\left[\frac{u(\tilde{\nu}_b,-\nu_b)^*u(\nu_b,-\nu_b)h_E(\nu_b)^*h_E(\tilde{\nu_b})}{\nu_b^*-\tilde{\nu}_b}\right] }.
\end{equation}
An example of such a bound state is shown in Fig. 1(c) of the main text.

To find the output after $N$ atoms for a given incident state $\vert \textup{in} \rangle$, we decompose the incident state into the eigenstates of the $S$-matrix, and then apply the $S$-matrix $N$ times, which just amounts to multiplying by the $N$th power of the eigenvalue in the decomposition \cite{Mahmoodian_2018}. In this way we find the real space representation [see also Eq. (3) in the main text]
\begin{equation}
	\psi_{\textup{out}}(x,x') = \langle x,x' \vert \textup{out} \rangle = \int dE' \left\{ \lambda_{E'}(\nu_b)^N\langle x,x' \vert E',\nu_b\rangle \langle E',\nu_b \vert \textup{in}\rangle + \int_{0}^{\infty} d\nu \lambda_{E'}(\nu)^N\langle x,x' \vert E',\nu\rangle \langle E',\nu \vert \textup{in}\rangle   \right\}, \label{eq:decompose-in}
\end{equation}
where $\vert E',\nu\rangle$ are the eigenstates of the $S$-matrix with real space representations given by Eq. \eqref{eq:eig-realspace}. Note that the integration  over $\nu$ in Eq. \eqref{eq:decompose-in} only ranges from $0$ to $\infty$ because changing the sign of $\nu$ gives the same eigenstate.

We take care of the $E'$-integration in Eq. \eqref{eq:decompose-in} using $\langle E',\nu | \textup{in}\rangle \propto \delta(E'-E)$, where $E$ is the total energy of the incoming photons, and handle the $\nu$-integration numerically. We consider only the case where the two incident photons have identical energies $\omega = E/2$, such that the input is $\langle x,x' \vert \textup{in} \rangle = e^{iE r_c}\sqrt{2}/(2\pi)$. The final expression for the wave function of the output is
\begin{equation}
	\begin{split}
		\psi_{\textup{out}}(x,x') &= 2\sqrt{2}ie^{iEr_c} \bigg\{\frac{\lambda_E(\nu_b)^N}{\xi_b(E,\nu_b)}\left[\frac{u_E(\nu_b,-\nu_b)h_E(\tilde{\nu}_b)}{\tilde{\nu}_b} - \frac{u_E(\tilde{\nu}_b,-\nu_b)h_E(\nu_b)}{\nu_b}\right]^* F_{E,\nu_b}(r) \\
			&\quad + \int_{\mathcal{D}_{\tilde{\nu}^2 < 0}} d\nu \frac{\lambda_E(\nu)^N}{\xi(E,\nu)}\left[\frac{u_E(\nu,\tilde{\nu}) - u_E(\tilde{\nu},-\nu)}{\nu}h_E(\nu) + \frac{u_E(\nu,-\nu)}{\tilde{\nu}}h_E(\tilde{\nu})\right]^*F_{E,\nu}^{(D = 0)}(r)  \\
			&\quad + \frac{1}{2}\int_{\mathcal{D}_{\tilde{\nu}^2 \geq 0}} d\nu \frac{\lambda_E(\nu)^N}{\xi(E,\nu)}\bigg\{\left[\frac{u_E(\nu,\tilde{\nu}) - u_E(\tilde{\nu},-\nu)}{\nu}h_E(\nu) + \frac{u_E(\nu,-\nu)}{\tilde{\nu}}h_E(\tilde{\nu})\right]^*F_{E,\nu}^{(D = 0)}(r) \\
			&\quad  + \left[\frac{u_E(\nu,-\tilde{\nu}) - u_E(-\tilde{\nu},-\nu)}{\nu}h_E(\nu) - \frac{u_E(\nu,-\nu)}{\tilde{\nu}}h_E(\tilde{\nu})\right]^*F_{E,\nu}^{(C = 0)}(r)\bigg\} \bigg\}, \label{eq:out-final}
	\end{split}
\end{equation}
where $\mathcal{D}_{\tilde{\nu}^2 < 0}$ is set of real-valued $\nu$ such that $\tilde{\nu}^2 < 0$, and $\mathcal{D}_{\tilde{\nu}^2 \geq 0}$ is set of real-valued $\nu$ such that $\tilde{\nu}^2 \geq 0$. In writing Eq. \eqref{eq:out-final} we partially used Eq. \eqref{eq:F_Enu_I} and partially Eq. \eqref{eq:F_Enu_II} for the decomposition of the input. In Figs. 2(b)-(d) of the main text we plot the wave function of the output, which is given by Eq. \eqref{eq:out-final}. The in-state in Fig. 2(a) of the main text is found by setting $N = 0$ in Eq. \eqref{eq:out-final}, which means that Fig. 2(a) of the main text also serves as a numerical completeness check of the eigenstates, since we get the expected result for the incident wave function.

We note that $|\psi_{\textup{out}}(x,x')|^2$ yields the second order correlation function of the transmitted photons output up to normalization
\begin{equation}
g^{(2)}(r) = 2\pi^2|\psi_{\textup{out}}(x,x')|^2
\end{equation}
that is discussed in the main text and shown in Fig.~1 of the main text.

In summary, we analyzed three different classes of $S$-matrix eigenstates: Continuum eigenstates that consist only of plane waves, continuum eigenstates consisting of plane waves and a localized term, and a bound state. The general expression for these states are given by Eq. \eqref{eq:eig-realspace}, where $F_{E,\nu}(r)$ is given by Eqs. \eqref{eq:F_Enu_I} and \eqref{eq:F_Enu_II} for the continuum states and Eq. \eqref{eq:F_bound} for the bound state. These expressions are used to plot the eigenstates in Fig. 1(c) of the main text. 

\section{Photon propagation in the presence of dissipation}
The addition of dissipative photon loss into non-guided modes ($\gamma > 0$) renders the $S$-matrix non-unitary. It is therefore no longer guaranteed that its eigenstates with different eigenvalues are orthogonal which poses problems for the approach based on eigenstate decomposition. 

In this section, we therefore describe another method that treats the chain as a cascaded quantum system, to derive the effect of $N$ emitters from the underlying single-emitter scattering physics. To this end, we find the eigenstates of the full single-atom Hamiltonian \eqref{eq:Hsingle} to determine the outgoing wave function after the scattering off a single atom for a given incoming boundary condition \cite{zgb12,zgb10}. The outgoing wave function after the $j$th atom can then be used as the input for atom $j+1$ to find the output after $j+1$ atoms. This yields a recursion relations, which can be used to find the output after all $N$ atoms and thereby the two-photon correlation function \cite{zgb12,zgb10}.

We denote the positions of the atoms by $x_j$ for $j = 1,2,\dots,N$, and we, moreover, define $x_0 = -\infty$ and $x_{N+1} = +\infty$. The wave function of the eigenstate of the Hamiltonian is given by Eq. \eqref{eq:general-form}, where $F(r)$ is continuous when $x,x' \neq x_j$ for all $j = 1,2,\dots,N$. So $F(r)$ can be written as
\begin{equation}
	F(r) = F_{\ell,j}(r), \quad x_{\ell} < x < x_{\ell+1} \textup{ and } x_j < x' < x_{j+1},
\end{equation}
so $F_{\ell,j}(r)$ [together with Eq. \eqref{eq:general-form}] gives the wave function when one photon has passed $\ell$ atoms, and the other has passed $j$ atoms. Still we only consider the case where $r = x - x' < 0$, which implies that $\ell \leq j$ in what follows.  We aim to find $F_{N,N}(r)$, which gives the output of the entire chain through
\begin{equation}
	\psi_{\textup{out}}(x,x') = e^{iEr_c}F_{N,N}(r), \label{eq:output-general}
\end{equation} 
where $E$ is the energy. 

As previously, we restrict the input to be two photons with identical energies $\omega = E/2$, which formally means that $F_{0,0}(r) = T_0$, where $T_0$ is a constant that defines the normalization. It is assumed that output after both photons have passed $j$ atoms  has the form
\begin{equation}
	F_{j,j}(r) = T_j + \sum_{n=0}^{j-1}\left[A_{j,n}e^{\kappa_1 r} + B_{j,n}e^{\kappa_2 r}\right]\frac{r^n}{n!}, \label{eq:Fjj}
\end{equation}
where $T_j$, $A_{j,n}$, and $B_{j,n}$ are constants to be determined, and $\kappa_1$, $\kappa_2$ are the solutions of
\begin{equation}
	\kappa^2 - \eta\kappa + \alpha = 0. \label{eq:kappa}
\end{equation}
It can be shown that the real parts of $\kappa_1$ and $\kappa_2$ are positive, so $F_{j,j}(r)$ converges for $r\to -\infty$. The assumption that $F_{j,j}(r)$ is of the form \eqref{eq:Fjj} will be justified by showing that $F_{j+1,j+1}(r)$ is of the form \eqref{eq:Fjj} given that $F_{j,j}(r)$ is. Note that $F_{0,0}(r)$ is of the form \eqref{eq:Fjj}. As mentioned above, we aim to find $F_{N,N}(r)$ because this gives the output after all $N$ atoms. We also note that if we set $T_0 = 1$, then $T_N$ is the transmission coefficient for two uncorrelated (individual) photons, since $F_{N,N}(r\to -\infty) = T_N$. The two-photon correlation function of the output (still with $T_0 = 1$) is $g^{(2)}(r) = |F_{N,N}(r)|^2/|T_N|^2$, so finding $F_{N,N}(r)$ directly gives the two-photon correlation function, which we plot in Fig. 3(a) of the main text.

Now that it has been established that $F_{N,N}(r)$ contains the information we search for, we derive a recursive relation for $F_{j,j}(x)$. To this end, we use Eqs. \eqref{eq:F1} and \eqref{eq:F2} that now read
\begin{align}
	\left(\partial_r^2 + \eta\partial_r + \alpha\right)F_{j,j+1}(r) &= \left(\partial_r^2 + \eta\partial_r + \alpha\right)F_{j,j}(r) + \left(\mu - \Gamma\partial_r\right)F_{j,j}(r), \label{eq:pass-first} \\
	\left(\partial_r^2 - \eta\partial_r + \alpha\right)F_{j+1,j+1}(r) &= \left(\partial_r^2 - \eta\partial_r + \alpha\right)F_{j,j+1}(r) + \left(\mu + \Gamma\partial_r\right)F_{j,j+1}(r). \label{eq:pass-second}
\end{align}
To find $F_{j,j+1}$ from $F_{j,j}$ we start by substituting Eq. \eqref{eq:Fjj} into Eq. \eqref{eq:pass-first} using Eq. \eqref{eq:kappa}
\begin{align}
	\MoveEqLeft
	\left(\partial_r^2 + \eta\partial_r + \alpha\right)F_{j,j+1}(r) \nonumber \\
		&= (\alpha+\mu)T_j + \sum_{n=0}^{j-1}\left[ ((2\eta-\Gamma)\kappa_1 + \mu)A_{j,n} e^{\kappa_1 r} + ((2\eta-\Gamma)\kappa_2 + \mu)B_{j,n} e^{\kappa_2 r} \right]\frac{r^n}{n!} \nonumber \\
			&\quad + \sum_{n=1}^{j-1}\left[(2\kappa_1 + \eta - \Gamma)A_{j,n}e^{\kappa_1 r} + (2\kappa_2 + \eta - \Gamma)B_{j,n}e^{\kappa_2 r}\right]\frac{r^{n-1}}{(n-1)!} + \sum_{n=2}^{j-1}\left[A_{j,n}e^{\kappa_1 r} + B_{j,n}e^{\kappa_2 r}\right]\frac{r^{n-2}}{(n-2)!}. \label{eq:pass-first-use-ansatz}
\end{align}
To find a solution of this equation we use the ansatz
\begin{equation}
	F_{j,j+1}(r) = \bar{T}_j + \sum_{n=0}^{j-1}\left[\bar{A}_{j,n}e^{\kappa_1 r} + \bar{B}_{j,n}e^{\kappa_2 r}\right]\frac{r^n}{n!}. \label{eq:Fjj+1}
\end{equation}
By substituting this ansatz into Eq. \eqref{eq:pass-first-use-ansatz} and using Eq. \eqref{eq:kappa} we obtain
\begin{equation}
	\bar{T}_j = \frac{\alpha + \mu}{\alpha}T_j, \label{eq:cjj+1}
\end{equation}
and
\begin{subequations}
	\label{eq:ABbar}
	\begin{align}
	2\eta\kappa_1\bar{A}_{j,j-1} &= [(2\eta-\Gamma)\kappa_1 + \mu]A_{j,j-1}, \\
	2\eta\kappa_1\bar{A}_{j,j-2} + (2\kappa_1 + \eta)\bar{A}_{j,j-1} &= [(2\eta-\Gamma)\kappa_1 + \mu]A_{j,j-2} + (2\kappa_1 + \eta - \Gamma)A_{j,j-1}, \\
	2\eta\kappa_1\bar{A}_{j,n} + (2\kappa_1 + \eta)\bar{A}_{j,n+1} + \bar{A}_{j,n+2} &= [(2\eta-\Gamma)\kappa_1 + \mu]A_{j,n} + (2\kappa_1 + \eta - \Gamma)A_{j,n+1} + A_{j,n+2}, \quad 0\leq n \leq j-3, \\
	2\eta\kappa_2\bar{B}_{j,j-1} &= [(2\eta-\Gamma)\kappa_2 + \mu]B_{j,j-1}, \\
	2\eta\kappa_2\bar{B}_{j,j-2} + (2\kappa_2 + \eta)\bar{B}_{j,j-1} &= [(2\eta-\Gamma)\kappa_2 + \mu]B_{j,j-2} + (2\kappa_2 + \eta - \Gamma)B_{j,j-1}, \\
	2\eta\kappa_2\bar{B}_{j,n} + (2\kappa_2 + \eta)\bar{B}_{j,n+1} + \bar{B}_{j,n+2} &= [(2\eta-\Gamma)\kappa_2 + \mu]B_{j,n} + (2\kappa_2 + \eta - \Gamma)B_{j,n+1} + B_{j,n+2},\quad 0\leq n \leq j-3,
	\end{align}
\end{subequations}
which specify all the constants in Eq. \eqref{eq:Fjj+1}. A general solution of Eq. \eqref{eq:pass-first-use-ansatz} can be found by adding a solution of the homogeneous system
\begin{equation}
	\left(\partial_r^2 + \eta\partial_r + \alpha\right)f_{j,j}(r) = 0 \label{eq:pass-first-homogeneous}
\end{equation}
to Eq. \eqref{eq:Fjj+1}, but as all non-vanishing solutions of Eq. \eqref{eq:pass-first-homogeneous} diverges for $r \to -\infty$, we find that Eq. \eqref{eq:Fjj+1} is the physical solution.

To find $F_{j+1,j+1}(r)$ we now substitute Eq. \eqref{eq:Fjj+1} into Eq. \eqref{eq:pass-second} using Eq. \eqref{eq:kappa}
\begin{align}
	\MoveEqLeft
	\left(\partial_r^2 - \eta\partial_r + \alpha\right)F_{j+1,j+1}(r) \nonumber\\
	&= (\alpha + \mu)\bar{T}_j  + \sum_{n=0}^{j-1}\left[(\mu+\Gamma\kappa_1)\bar{A}_{j,n}e^{\kappa_1 r} + (\mu + \Gamma\kappa_2)\bar{B}_{j,n}e^{\kappa_2 r} \right]\frac{r^n}{n!} \nonumber\\
		&\quad+ \sum_{n=1}^{j-1}\left[(2\kappa_1 - \eta + \Gamma)\bar{A}_{j,n}e^{\kappa_1 r} + (2\kappa_2-\eta+\Gamma)\bar{B}_{j,n}e^{\kappa_2 r} \right]\frac{r^{n-1}}{(n-1)!} +  \sum_{n=2}^{j-1}\left[\bar{A}_{j,n}e^{\kappa_1 r} + \bar{B}_{j,n}e^{\kappa_2 r} \right]\frac{r^{n-2}}{(n-2)!}.
\end{align}
We solve this differential equation using the ansatz
\begin{equation}
F_{j+1,j+1}(r) = T_{j+1} + \sum_{n=0}^{j}\left[A_{j+1,n}e^{\kappa_1 r} + B_{j+1,n}e^{\kappa_2 r}\right]\frac{r^n}{n!} \label{eq:Fj+1j+1}
\end{equation}
and Eqs. \eqref{eq:kappa} and \eqref{eq:cjj+1}. We find
\begin{equation}
	T_{j+1} = \frac{(\alpha + \mu)^2}{\alpha^2}T_j \label{eq:cj+1j+1},
\end{equation}
and 
\begin{subequations}
	\label{eq:AB}
	\begin{align}
		(2\kappa_1 - \eta)A_{j+1,j} &= (\mu + \Gamma\kappa_1)\bar{A}_{j,j-1}, \\
		(2\kappa_1 - \eta)A_{j+1,j-1} + A_{j+1,j} &= (\mu + \Gamma\kappa_1)\bar{A}_{j,j-2} + (2\kappa_1 - \eta + \Gamma)\bar{A}_{j,j-1}, \\
		(2\kappa_1 - \eta)A_{j+1,n} + A_{j+1,n+1} &= (\mu + \Gamma\kappa_1)\bar{A}_{j,n-1} + (2\kappa_1 - \eta + \Gamma)\bar{A}_{j,n} + \bar{A}_{j,n+1}, \quad 1\leq n \leq j-2, \\
		(2\kappa_2 - \eta)B_{j+1,j} &= (\mu + \Gamma\kappa_2)\bar{B}_{j,j-1}, \\
		(2\kappa_2 - \eta)B_{j+1,j-1} + B_{j+1,j} &= (\mu + \Gamma\kappa_2)\bar{B}_{j,j-2} + (2\kappa_2 - \eta + \Gamma)\bar{B}_{j,j-1}, \\
		(2\kappa_2 - \eta)B_{j+1,n} + B_{j+1,n+1} &= (\mu + \Gamma\kappa_2)\bar{B}_{j,n-1} + (2\kappa_2 - \eta + \Gamma)\bar{B}_{j,n} + \bar{B}_{j,n+1}, \quad 1\leq n \leq j-2.
	\end{align}
\end{subequations}
These equations specify all the constants in Eq. \eqref{eq:Fj+1j+1} except $A_{j+1,0}$ and $B_{j+1,0}$. This is because $f_{j+1,j+1}(r) = A_{j+1,0}e^{\kappa_1 r} + B_{j+1,0}e^{\kappa_2 r}$ is [by Eq. \eqref{eq:kappa}] the full solution of the homogeneous system
\begin{equation}
	\left(\partial_r^2 - \eta\partial_r + \alpha\right)f_{j+1,j+1}(r) = 0.
\end{equation}
Hence the constants $A_{j+1,0}$ and $B_{j+1,0}$ must be determined by the boundary conditions, i.e. by Eqs. \eqref{eq:boundary-F}, which yield
\begin{subequations}
	\label{eq:ABj+1,0}
	\begin{align}
		A_{j+1,0} + B_{j+1,0} &= -\frac{\mu^2}{\alpha^2}T_j + 2(\bar{A}_{j,0} + \bar{B}_{j,0}) - A_{j,0} - B_{j,0}, \\
		\kappa_1 A_{j+1,0} + \kappa_2 B_{j+1,0} &= A_{j,1} + B_{j,1} - A_{j+1,1} - B_{j+1,1} + (\kappa_1-\Gamma)A_{j,0} + (\kappa_2 - \Gamma)B_{j,0} + \Gamma(\bar{A}_{j,0} + \bar{B}_{j,0}) + \Gamma\frac{\mu}{\alpha}T_j.
	\end{align}
\end{subequations}
Since $F_{j+1,j+1}(r)$, given by Eq. \eqref{eq:Fj+1j+1}, is indeed of the form \eqref{eq:Fjj}, Eqs. \eqref{eq:Fjj}, \eqref{eq:ABbar}, \eqref{eq:cj+1j+1}, \eqref{eq:AB}, and \eqref{eq:ABj+1,0} give a recursive algorithm for $F_{j,j}(r)$. Applying this algorithm successively gives $F_{N,N}(r)$, which gives the output after the full chain.

As noted above, $T_N$ is the transmission coefficient for two uncorrelated photons if we set $T_0 = 1$. The probability that two uncorrelated photons will be transmitted through the chain is then given by $|T_N|^2$, and by Eq. \eqref{eq:cj+1j+1} we have
\begin{equation}
	|T_N|^2 = \left|\frac{\alpha + \mu}{\alpha}\right|^{4N}. \label{eq:TNsq}
\end{equation}
Inserting the definitions of $\alpha$ and $\mu$ [Eqs. \eqref{eq:alpha} and \eqref{eq:mu}] into Eq. \eqref{eq:TNsq} gives Eq. (4) of the main text. The corresponding two-photon correlation function is given by $g^{(2)}(r) = |F_{N,N}(r)|^2/|T_N|^2$, which we show in Fig. 3(a) of the main text.

\section{Optimal atom number}
\begin{figure}
	\includegraphics{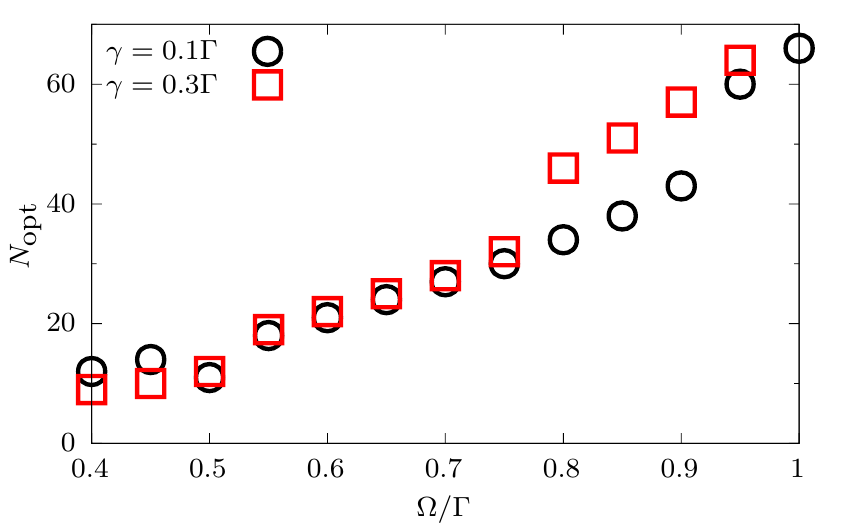}
	\caption{Minimal atom number $N_{\textup{opt}}$ needed to obtain $g^{(2)}(0) \leq 0.1$ as a function of the Rabi frequency  $\Omega$ of the coupling field. The detunings are $\Delta = 0.4\Gamma$ and $\bar{\Delta} = -0.2\Gamma$. \label{fig:Nmin}}
\end{figure}

In the main text we describe that the optimal atom number $N_{\textup{opt}}$ to obtain $g^{(2)}(0) \leq 0.1$ increases as a function of the Rabi frequency $\Omega$ in the regime of large $\Omega$. This can be seen in Fig. \ref{fig:Nmin}. The values of $N_{\textup{opt}}$ shown in Fig. \ref{fig:Nmin} are the values underlying the transmission displayed in Fig. 3(c) of the main text.
\twocolumngrid

\end{document}